\documentclass[twocolumn, showpacs, prb]{revtex4-1}
\usepackage{graphicx, color}
\def\ind#1{{_{\mathrm{#1}}}}

\begin{document}

\title{Influence of the magnetoelectric coupling on the electric field induced magnetization reversal in a composite  {non-strained} multiferroic chain}

\author{Paul P. Horley$^1$, Alexander Sukhov$^2$, Chenglong Jia$^2$, Eduardo Mart\'{\i}nez$^1$ and Jamal Berakdar$^2$}

\address{$^1$Centro de Investigaci\'{o}n en Materiales Avanzados (CIMAV S.C.), Chihuahua/Monterrey, 31109 Chihuahua, Mexico \\
$^2$Institut f\"ur Physik, Martin-Luther Universit\"at Halle-Wittenberg, 06120 Halle (Saale), Germany}

\date{\today}

\begin{abstract}
We study theoretically the multiferroic dynamics in a composite one-dimensional system
  consisting of  BaTiO$_3$   multiferroically coupled to an iron chain.
 The method treats the magnetization and the polarization as thermodynamic quantities describable via a combination of the Landau-Lifshits-Gilbert  and the Ginzburg-Landau dynamics coupled via   an additional term in the total free energy density. This term stems from the multiferroic interaction at the interface.  For a wide range of strengths of this coupling  we predict  the possibility of obtaining a well-developed hysteresis in the ferromagnetic part of the system induced by an external electric field. The dependence of the reversal modes on the electric field frequency is also investigated and we  predict a considerable stability of the magnetization reversal for frequencies in the range of $0.5-12$ [GHz].
\end{abstract}



\maketitle

\section{Introduction}
Multiferroics (MFs), i.e. materials that possess   ferromagnetic and ferroelectric properties,
 have recently attracted significant research
\cite{Fi05,EeMa06,RaSp07} as they hold the promise of qualitatively new device concepts such
 as electric field induced  magnetization switching  at low heat
dissipation.
In addition to single-phase MFs \cite{Ri09},  i.e.  materials with ferroelectric (FE) and ferromagnetic (FM) or antiferromagnetic (AFM) ordering (e.g.  BiFeO$_3$ \cite{WaNe03}),
 composite MFs \cite{NaBi08} are in the focus of current research. These are systems realized
 as heterostructures of a wide range of different materials \cite{Valencia11}. This diversity is hoped to  compensates for the scarcity of single-phase MFs for room-temperature applications and to offer
 new routs to the engineering of the magnetoelectric coupling and its enhancement,
  e.g. via an appropriate multilayer stacking.
 A typical example for such composite FE/FM junctions is BaTiO$_3$/Fe that was predicted \cite{DuJa06} and studied \cite{FeMa10} theoretically. Experimentally  BaTiO$_3$/Fe  was successfully realized and characterized. \cite{SaPo07,Meyerheim11} In addition, a recent publication reported on a considerable coupling between the iron and the barium titanate which renders possible  a change of the magnetization of the iron layer when  an electric field is applied, even at a room temperature.\cite{Valencia11} Another important issue concerns the significant variation of the tunnel magnetoresistance depending on the polarization direction of BaTiO$_3$ layer. \cite{Garcia10,Valencia11} All these experimental findings are very promising steps towards  the creation of memory elements \cite{BiBa08} based on composite MFs, which could be written by electric field pulses. Problems appearing on the way  to achieve this goal include the optimization of the geometry of FE/FM layers and the improvement of the interaction between them. This paper is dedicated to the latter issue.

In contrast to our earlier studies,\cite{SuJi10,JiSu11,Su11_2} where we focused on a more simple tetragonal phase of barium titanate in which the perovskite exists at room temperature,\cite{Me49} here we consider the rhombohedral phase \cite{Me49} at zero Kelvin with the corresponding ferroelectric potential.  {Under these circumstances it is possible to neglect the thermal stochastic effects on the effective fields. To improve the accuracy of the model, we also included the long-range interactions for both the FE and the FM materials and the effect of the FE depolarizing field.}

\section{Theoretical formulation}

We consider a chain composed of FE (BaTiO$_3$) and FM (Fe) sites as shown in Fig. 1.  The aim is to describe
the evolution of the magnetization (polarization) under the influence of
external fields.  We adopt a Ginzburg-Landau phenomenology based on  coarse grained order parameters that
formally  result from an averaging of the relevant microscopic quantities over an appropriate cell.
These cells are in our calculations for the whole chain
cubes (called hereafter sites) of equal volume $a^3_{\mathrm{FE}}=a^3_{\mathrm{FM}}=a^3=5\times5\times5$ [nm$^3$].
We will show results for a chain formed by five FE and five FM sites.
Calculations   performed for longer chains \cite{JiSu11}  showed that the total magnetization reversal is hardly achievable for a chain that has   more than ten magnetic sites.

The  coarse grained total free energy \cite{SuJi10,JiSu11} reads
\begin{equation}
F\ind{TOT}=a^3F_{\mathrm{FE}}+a^3F_{\mathrm{FM}}+E_{\mathrm{CON}}.
\label{eq_0}
\end{equation}
The free energy density of the ferroelectric
\begin{equation}
\displaystyle F\ind{FE} = F\ind{GLD} + F\ind{DEP} + F_{\mathrm{CPL}}^{\mathrm{FE}} + F_{\mathrm{DDI}}^{\mathrm{FE}} + F\ind{EXT}
\label{eq_1}
\end{equation}
includes
the  Ginzburg-Landau-Devonshire term $F\ind{GLD}$ \cite{Gi49,De49}
\begin{eqnarray}
&& \displaystyle F\ind{GLD}=\sum_i \left[ \alpha\ind{FE}_1(P_{i\mathrm{x}}^2+P_{i\mathrm{y}}^2+P_{i\mathrm{z}}^2) \right. \nonumber \\
&&+ \beta\ind{FE}_1(P_{i\mathrm{x}}^4+P_{i\mathrm{y}}^4+P_{i\mathrm{z}}^4) + \gamma\ind{FE}_{1}(P_{i\mathrm{x}}^6 + P_{i\mathrm{y}}^6 + P_{i\mathrm{z}}^6) \nonumber \\
&&+ \beta\ind{FE}_{2}(P_{i\mathrm{x}}^2 P_{i\mathrm{y}}^2 + P_{i\mathrm{y}}^2 P_{i\mathrm{z}}^2 + P_{i\mathrm{x}}^2 P_{i\mathrm{z}}^2) + \gamma\ind{FE}_{3}P_{i\mathrm{x}}^2 P_{i\mathrm{y}}^2 P_{i\mathrm{z}}^2 \nonumber \\
&& \left. + \gamma\ind{FE}_{2}\left(P_{i\mathrm{x}}^4 (P_{i\mathrm{y}}^2 + P_{i\mathrm{z}}^2) + P_{i\mathrm{y}}^4 (P_{i\mathrm{x}}^2 + P_{i\mathrm{z}}^2) + P_{i\mathrm{z}}^4 (P_{i\mathrm{x}}^2 + P_{i\mathrm{y}}^2) \right) \right] \nonumber \\
\label{eq_2}
\end{eqnarray}

with the expansion coefficients $\alpha\ind{FE}_1, \beta\ind{FE}_{1, 2}$ and $\gamma\ind{FE}_{1-3}$.

The  depolarizing energy density $F\ind{DEP}$ \cite{RaAh07} (cf. Fig. 1) reads
\begin{equation}
F\ind{DEP} = \frac{1}{2} \sum_i(\frac{2 \lambda\ind{M}}{a N} \frac{P_{i\mathrm{x}} P_{i\mathrm{x}}}{{\color{black}\varepsilon\ind{DL}\varepsilon_0}})
\label{eq_3}
\end{equation}
which involves the dead layer permittivity $\varepsilon\ind{DL}$, the dielectric constant in vacuum $\varepsilon_0$, the metal screening length $\lambda\ind{M}$, the FE cell size $a$, and the cell number $N$.
The ferroelectric nearest neighbors coupling energy density $F_{\mathrm{CPL}}^{\mathrm{FE}}$  is\cite{Hl06}
\begin{eqnarray}
F_{\mathrm{CPL}}^{\mathrm{FE}} = \kappa\ind{FE} \sum_i \left[ (P_{(i+1)\mathrm{x}}-P_{(i)\mathrm{x}})^2 \right. \nonumber \\
\left. + (P_{(i+1)\mathrm{y}}-P_{(i)\mathrm{y}})^2 + (P_{(i+1)\mathrm{z}}-P_{(i)\mathrm{z}})^2 \right] \nonumber
\label{eq_4}
\end{eqnarray}
where  $\kappa\ind{FE}$ is the ferroelectric coupling constant.
The  dipole-dipole interactions $F_{\mathrm{DDI}}^{\mathrm{FE}}$ is
\begin{equation}
F_{\mathrm{DDI}}^{\mathrm{FE}} = \frac{1}{4 \pi \varepsilon\ind{FE}\varepsilon_0}\sum_{i \neq k}\left[ \frac{\vec{P}_i \cdot \vec{P}_k - 3(\vec{P}_i \cdot \vec{e}_{ik})(\vec{e}_{ik} \cdot \vec{P}_k)}{n_{ik}^3} \right].
\label{eq_5}
\end{equation}
Here   $\varepsilon\ind{FE}$ is the ferroelectric permittivity, $\vec{e}_{ik}$
is a unit vector parallel to the line joining the centers of the  dipoles $\vec{P}_i$ and  $\vec{P}_k$
and $n_{ik}$ is the distance  (measured in   units of $a$) between the two dipoles.

The  energy term  $F\ind{EXT}$  stems from  the applied electric field $\vec{E}$
\begin{equation}
F\ind{EXT} = - \sum_i \vec{E} \cdot \vec{P}_i.
\label{eq_6}
\end{equation}

Analogously the ferromagnetic part is characterized by the coarse grained free energy density
\begin{equation}
F\ind{FM} = F\ind{ANI} + F_{\mathrm{XCG}}^{\mathrm{FM}} + F_{\mathrm{DDI}}^{\mathrm{FM}}
\label{eq_7}
\end{equation}
 which consists of the
(uniaxial) magnetocrystalline anisotropy contribution \cite{Ch02}
\begin{equation}
\displaystyle F\ind{ANI} = -\sum_j \frac{K_1}{M_{\mathrm{S}}^2}M_{x j}^2
\label{eq_8}
\end{equation}
with the anisotropy constant $K_1$ and the saturation magnetization $M_S$.
The nearest-neighbor exchange interaction has the form \cite{Ch02}
\begin{equation}
F_{\mathrm{XCG}}^{\mathrm{FM}} = -\sum_j\frac{A}{a^2 M_S^2}\vec{M}_j \cdot \vec{M}_{j+1}
\label{eq_9}
\end{equation}
where $A$ is the interaction constant,
and the dipole-dipole interaction is
\begin{equation}
F_{\mathrm{DDI}}^{\mathrm{FM}} = \frac{\mu_0}{4 \pi}\sum_{j \neq l}\left[ \frac{\vec{M}_j \cdot \vec{M}_l - 3(\vec{M}_j \cdot \vec{e}_{jl})(\vec{e}_{jl} \cdot \vec{M}_l)}{n_{jl}^3} \right]
\label{eq_10}
\end{equation}
where  $\mu_0$ is the susceptibility constant.
The coupling between the ferroelectric and the ferromagnetic parts \cite{VaHo10} is due to the mobile spin-polarized electrons accumulated at the interface in order to screen the
electric polarization in the FE part.\cite{CaJu09} A change of the accumulated spin density (e.g. due to a change of the electric polarization) will act with a torque on the magnetization. This is however a surface effect restricted to the region in the vicinity of the interface (which is the reason why we are considering short chains).   In other words, only surface cells (those with index 1 in Fig. 1) will participate in this coupling \cite{LeSa10} which contributes with   the interaction energy
\begin{equation}
E\ind{CON} = a^3\lambda \vec{P}_1 \cdot \vec{M}_1.
\label{eq_11}
\end{equation}

The time dynamics of the polarization $\vec{P}_i$ and the magnetization $\vec{M}_j$ of the individual sites are obtained by propagating  the Landau-Khalatnikov (LKh) \cite{LaKh53,SiWi04} equation
\begin{equation}
\gamma_{\nu}\frac{d \vec{P}_i}{dt} = \vec{E}_{\mathrm{FE} i}\equiv -\frac{1}{a^3}\frac{\delta F\ind{TOT}}{\delta \vec{P}_i},
\label{eq_12}
\end{equation}

and the Landau-Lifshitz-Gilbert (LLG) \cite{LaLi35,Gi55} equation
\begin{eqnarray}
\frac{d\vec{M}_j}{dt}=-\frac{\gamma}{1+\alpha_{\mathrm{FM}}^2}\left\{[\vec{M}_j \times \vec{H}_{\mathrm{FM} j}] \right.\nonumber \\ \left.+ \frac{\alpha\ind{FM}}{M_S}\vec{M}_j(\vec{M}_j \cdot \vec{H}_{\mathrm{FM}_j}) - \alpha\ind{FM} M_S \vec{H}_{\mathrm{FM} j}\right\} .
\label{eq_13}
\end{eqnarray}

The coefficients entering into the LKh are the viscosity constant $\gamma_\nu$ and the effective electric field $\vec{E}_{\mathrm{FE} i}$. The  LLG equation involves  the gyromagnetic ratio $\gamma$, the Gilbert damping coefficient $\alpha\ind{FM}$, the saturation magnetization $M_S$, and the effective field $\displaystyle \vec{H}_{\mathrm{FM} j}=-\frac{1}{a^3}\frac{\delta F\ind{TOT}}{\delta \vec{M}_j}$.\\
 From a computational point of view the appropriate  choice of the cell size $a$ is important. If it is chosen too small, the coarse graining procedure to obtain the macroscopic quantities $\vec{P}_i$ and $\vec{M}_j$  becomes questionable and one faces in addition problems with the superpara-magnetic/electric limits. For a cell too large, a multi-domain state sets in.  The  cell  size used in our calculation
 is therefore $a = 5~[\mathrm{nm}]$ for both FE and FM parts of the chain.
 The material parameters were set as those of barium titanate at $T=0$~[K].
  Specifically, we choose $\alpha\ind{FE 1} = -1.275 \times 10^8~[\mathrm{V \cdot m/C}]$,\cite{Appendix} $\beta\ind{FE 1} = -2.045 \times 10^9~[\mathrm{V \cdot m^5/C^3}]$,\cite{Appendix} $\beta\ind{FE 2} = 3.230 \times 10^8~[\mathrm{V \cdot m^5/C^3}]$,\cite{Appendix} $\gamma\ind{FE 1} = 9.384 \times 10^9~[\mathrm{V \cdot m^9/C^5}]$,\cite{Appendix} $\gamma\ind{FE 2} = 4.470 \times 10^9~[\mathrm{V \cdot m^9/C^5}]$,\cite{Appendix} $\gamma\ind{FE 3} = 4.919 \times 10^9~[\mathrm{V \cdot m^9/C^5}]$,\cite{Appendix} $\gamma\ind{\nu} = 1.5 \times 10^{-5}~[\mathrm{V \cdot m \cdot s/C}]$,\cite{Hl07} $P\ind{S} = 0.499~[\mathrm{C/m^2}]$,\cite{calc_P}$ \kappa\ind{FE}=2.04 \times 10^8~[\mathrm{V \cdot m/C}]$,\cite{Hl06} $\varepsilon\ind{FE} = 164.$\cite{calc_epsilon} The depolarizing energy density $F\ind{DEP}$ includes the permittivity of the metallic electrodes, or the so called ``dead layer",\cite{BrLe09} which can be lower than the permittivity of the material. This parameter is difficult to measure \cite{KiJo05}.
  In the numerical calculation we used the reasonable value $\varepsilon\ind{DL}$ = $\varepsilon\ind{FE}/2 = 82$. The material of the FM layer is taken as iron with the  following parameters at $T=0$~[K]: $\alpha\ind{FM}= 0.5$,\cite{FM_damping} $K_1 = 4.8 \times 10^4~[\mathrm{J/m}]$,\cite{Coey10} $M\ind{S} = 1.71 \times 10^6~[\mathrm{A/m}]$,\cite{Coey10} $A = 2.1 \times 10^{-11}~[\mathrm{J/m}]$.\cite{Coey10}

A special role in the multiferroic dynamics is played by the  coupling strength $\lambda$. We estimate the strength of the magnetoelectric coupling parameter related to the BaTiO$_3$/Fe-interface using the proposed \textit{ab-initio} expression \cite{DuJa06,DuVe08} $\alpha\ind{S}=\mu_0 \Delta M/E\ind{C}$, where the surface magnetoelectric coupling $\alpha\ind{S}$ is defined as the change of the surface magnetization $\Delta M$ for the electric coercive field $E\ind{C}$. The coercive field can be estimated as $E\ind{C}\approx P\ind{s}/(\varepsilon\ind{FE}\varepsilon_0)$. Keeping in mind that the coupling energy (\ref{eq_11}) might also be expressed through the induced magnetization $\Delta M_0$ at the interface and the net ferromagnetic magnetization $M$ as $E\ind{CON}=-J/(M_{\mathrm{S}}^2) \Delta \vec{M}_1 \cdot \vec{M}$, we finally obtain $\lambda=J \alpha\ind{S}/(\mu_0 M_{\mathrm{S}}^2 a^4 \varepsilon\ind{FE}\varepsilon_0)$. For $\alpha\ind{S}=0.2\cdot 10^{-17}$~[T m$^2$/V] \cite{FeMa10} the last expression yields $\lambda \approx 0.063$~[s/F]. This numerical value did not result in any sizable interaction between FE and FM layers in the model considered. We attribute this situation to the difficulty in the numerical definition of $\lambda$, because the expression for the coupling constant  depends strongly on the size of the interface cell $a$. Under these circumstances, the most adequate approach was to consider $\lambda$ as a variable, with the aim to obtain the general picture of magnetization reversal induced by an electric field as a function of the magnetoelectric coupling.

\begin{figure}
\includegraphics[scale=1.0]{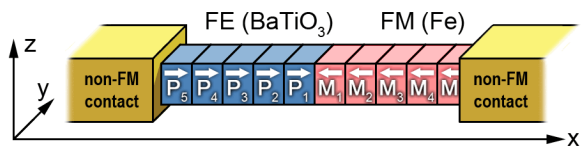}
\label{Fig1}
\caption{Schematics  of the composite multiferroic structure formed of five ferroelectric and five ferromagnetic sites. The initial state for both the FE polarization and the magnetization is chosen as random.}
\end{figure}

\begin{figure}
\includegraphics[scale=1.0]{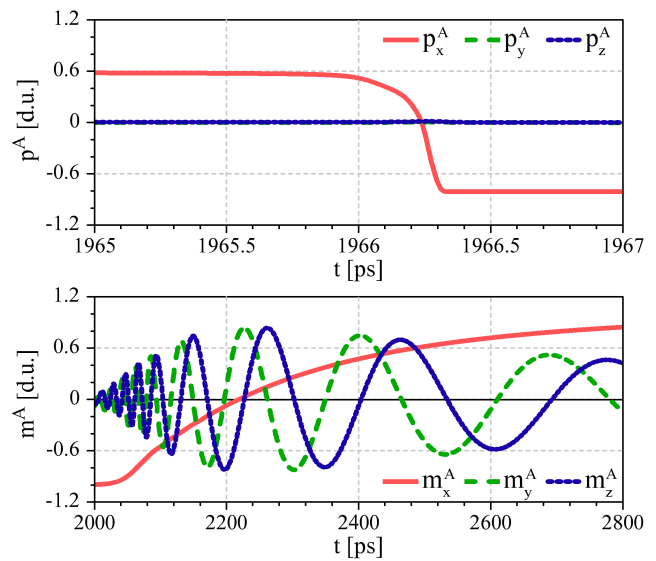}
\label{Fig2}
\caption{The difference in the time scale for the ferroelectric (upper panel) and the ferromagnetic (lower panel) reversals. The FE/FM coupling constant is taken as $\lambda = 10$ [s/F].}
\end{figure}

As in our previous studies \cite{SuJi10,JiSu11} the MF-chain will be characterized by the averaged total polarization $\vec{p}^{\mathrm{A}}=(N P\ind{S})^{-1}\sum_i \vec{P}_i$ and the averaged net magnetization $\vec{m}^{\mathrm{A}}=(N M\ind{S})^{-1}\sum_j \vec{M}_j$.

To reverse the FM part  $via$ the FE part of the chain, we apply the harmonic electric field $E_{\mathrm{x}}(t) = E_0 \sin (\omega t)$ with the  amplitude $E_0 = 8 \times 10^7$ [V/m]. As a frequency we choose at first $\omega/(2\pi) = 2$ [GHz]. The time profiles of the field reversal for the  ferroelectric and the ferromagnetic parts of the system are given in Fig. 2. As one can see, the ferroelectric part re-polarizes quickly within approximately 300 [fs]. The reversal modifies only the x-component of the polarization, with a very minor variation of the  z-component in the vicinity of the point where $p_{\mathrm{x}}^{\mathrm{A}}$ changes its sign. This happens because under a negative bias the polarization vector of the pre-contact diminishes in magnitude to zero, and then reverses the direction along the field. The remaining  cells flip their polarization vectors along the x-axis without a precession, causing a fast and a well-developed  reversal. On the contrary, the magnetization reversal requires a much larger time up to a nanosecond. As one can see from the plot, the re-orientation of the magnetization vector involves a heavy precession with considerable deviations of the $m_y^A$ and $m_z^A$ components from zero. These oscillations have a higher frequency at the beginning of the reversal process, gradually lowering the frequency and the amplitude till the equilibrium state is reached. Naturally, such a distinct time scale of the system components poses a significant problem for a proper modeling. To ensure the accurate numerical solution, one should keep the integration time step at femtoseconds, which drastically increases the number of integration steps needed to follow the system dynamics on time scales adequate to observe the hysteresis curves formed under the field variation with GHz frequencies. This large number of steps definitely will be an issue for the calculations of  large systems composed of hundreds of particles due to the need to evaluate the long range dipole-dipole interaction fields in the ferroelectric and the ferromagnetic parts. In our case, the LKh and the LLG equations were integrated with the Heun method using time step $\Delta t$ = 2 [fs].

\begin{figure*}
\includegraphics[scale=1.2]{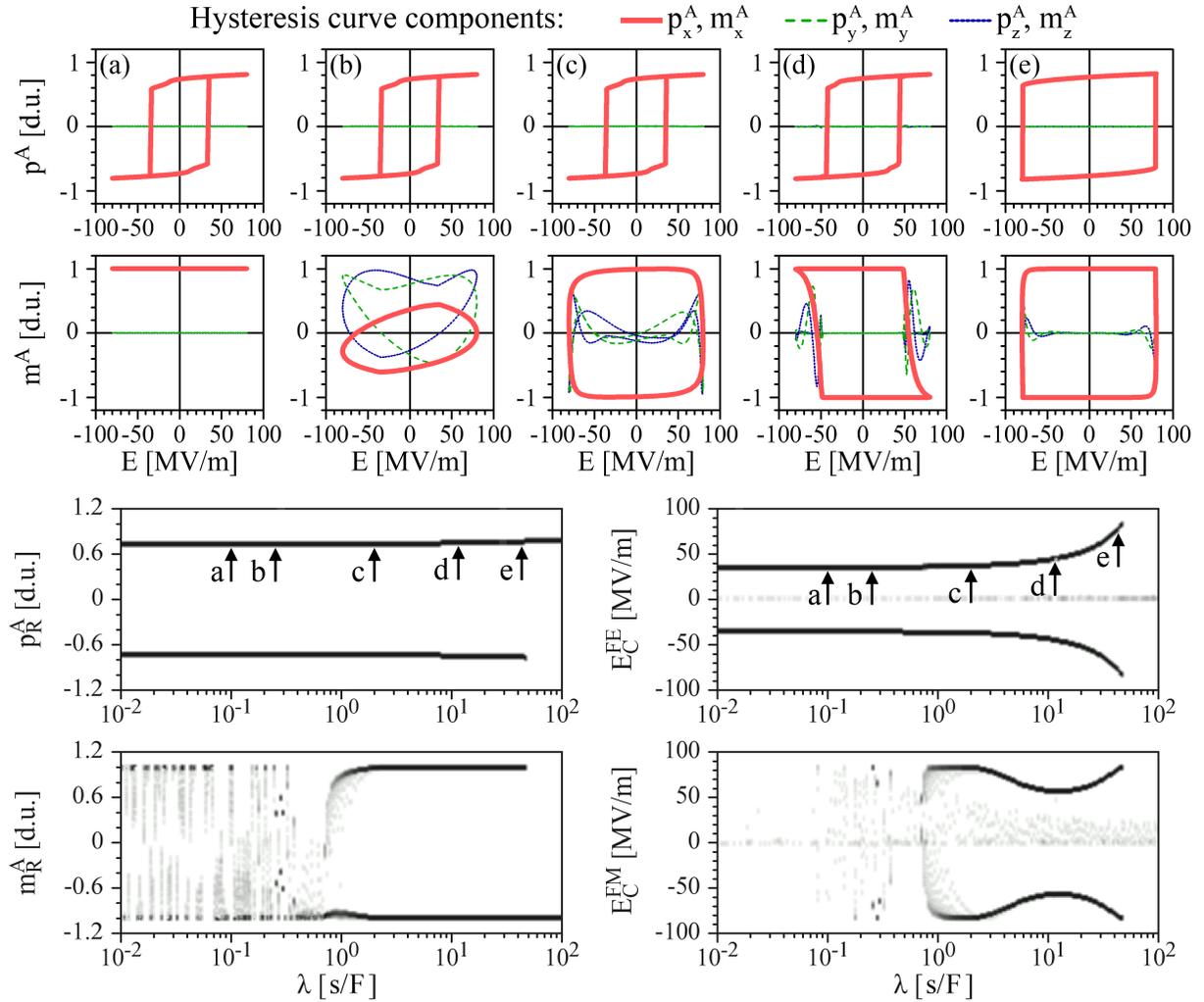}
\label{Fig3}
\caption{The dependence of the multiferroic reversal on the strength of FE/FM interaction $\lambda$. The hysteresis curves are presented for: a) $\lambda$ = 0.1 [s/F], b) $\lambda$ = 0.255 [s/F], c) $\lambda$ = 2 [s/F], d) $\lambda$ = 12 [s/F], e) $\lambda$ = 45.5 [s/F]. The hystograms in the lower part of the figure depict the remanence ($p^A_R$, $m^A_R$) end coercitivity ($E^{FE}_C$, $E^{FM}_C$) for the  averaged hysteresis curves obtained for ferroelectric and ferromagnetic parts of the chain. The applied electric field is characterized by the amplitude $E_0$ = 80 [MV/m] and frequency $\omega/(2\pi)$ = 2 [GHz].}
\end{figure*}

\section{Numerical results and discussions}

To study the influence of the coupling constant on the behavior of the system, it is necessary to select parameters that allow an easy and a reliable characterization of the hysteresis curve. We propose to use the values of the averaged polarization and the magnetization ($p^A$ and $m^A$, respectively) at a zero applied field, as well as the field values $E_C$ for which the magnetization or the  polarization are equal to zero. If the hysteresis loop is properly formed, these points will define the remanence and the coercitivity of the hysteresis loop. If the hysteresis is absent, it will become immediately notable, for example, by the existence of a single remanence point or the presence of various zeros of magnetization / polarization curve. To collect the data shown in Fig. 3 we integrated the system for 15 full field cycles and plotted the density of $p_{\mathrm{R}}^{\mathrm{A}} = p^{\mathrm{A}}(E=0)$, $m_{\mathrm{R}}^{\mathrm{A}} = m^{\mathrm{A}}(E=0)$ and $E_{\mathrm{C}}^{\mathrm{FE}}$, $E_{\mathrm{C}}^{\mathrm{FM}}$ as a function of the coupling parameter $\lambda$. These plots are shown in the bottom part of Fig. 3. This approach allows to see if the hysteresis curve is well-defined and repeatable, which will result in sharp and dark lines on the corresponding plot. If a hysteresis loop is unstable and varies from cycle to cycle, one  obtains the grayish bands instead of the thin dark curves. The light-colored noise for the positive fields on the coercitivity diagrams correspond to the points when the polarization / magnetization changes  sign before reaching the stable hysteresis curve. In some cases, such ``self-adjustment" of the phase trajectory takes place during the several  first cycles of  the electric field, which complicates the definition of an exact threshold that would allow to remove this noise. On the other hand, as we are studying steady hysteresis curves observed during the 15$^{th}$ field cycle, such initial noise is irrelevant and can be easily neglected. The most characteristic cases with FE and FM hysteresis curves illustrated in the upper part of the figure are marked with arrows in remanence and coercitivity diagrams.

As  seen from the figure  the coupling strength $\lambda < 0.8$ [s/F] is insufficient to reverse the magnetization. For the small value  $\lambda=0.1$ [s/F] (Fig. 3a) the value of $m_{\mathrm{x}}^{\mathrm{A}}$ is constant and matches   the saturation magnetization of the material, while the ferroelectric part manages to reach an entire hysteresis cycle. The steps at the edges of the hysteresis are caused by the discussed peculiarity of FE reversal mechanism that involves a vanishing of the first site before the field achieves the reversal value. For a stronger coupling $\lambda = 0.255$ [s/F] (Fig. 3b) the torque rendered by the polarization ``kicks" the magnetization with a frequency that is similar to the FM precession frequency, resulting in a periodic variation of $m_{\mathrm{x}}^{\mathrm{A}}$, $m_{\mathrm{y}}^{\mathrm{A}}$ and $m_{\mathrm{z}}^{\mathrm{A}}$. These oscillation modes may be interesting as a way to achieve a resonant magnetization precession. However, they do not allow to reach a saturation magnetization and to form the hysteresis loop, which is required  for the operation as memory devices.

The full ferromagnetic hysteresis loop emerges for $\lambda > 0.8$ [s/F], quickly reaching the full saturation value $M_S$. Due to the fact that the FM part is much slower than FE, the magnetic hysteresis has a larger coercitivity, however. This is most clearly seen for the case of a relatively weak coupling $\lambda = 2$ [s/F] (Fig. 3c), where FM dynamics definitely can not follow FE dynamics  at the desired speed. We observe pronounced oscillations in the $m_{\mathrm{y}}^{\mathrm{A}}$ and $m_{\mathrm{z}}^{\mathrm{A}}$ components upon the field-induced reversal. It is worth mentioning that the FE part of the system is not influenced much by the variation of the magnetization for these values of $\lambda$, for  the ferroelectric hysteresis exhibits  only a minor enhancement of the remanence and has almost the same coercitivity. An increase in the magneto-coupling strenght  $\lambda$ to 12 [s/F] (Fig. 3d) seems to yield  desirable results: the  magnetization hysteresis is saturated and narrow, with a fast relaxation of the $m_{\mathrm{y}}^{\mathrm{A}}$ and $m_{\mathrm{z}}^{\mathrm{A}}$ components after the full reversal of $m_{\mathrm{x}}^{\mathrm{A}}$. As the FE/FM feedback becomes more pronounced, the hysteresis of FE shows a larger coercitivity and remanence. The observed ``mirror-symmetry" of the FE and FM hysteresis loops has its origin in the opposite sign of the coupling terms in the LKh and LLG equations. For a large coupling $\lambda = 45.5$ [s/F] (Fig. 3e) both hysteresis curves deteriorate, as the ferromagnetic part starts to  ``hold"  the FE sites, hindering  their reversal until a considerable electric field is applied. A further increase of   the coupling strength   destroys the  hysteresis of both the  FE and FM layers as $\lambda$ exceeds 60 [s/F].

To address the frequency dependence of the composite multiferroic reversal, we calculated the  remanence and coercitivity diagrams (Fig. 4) for the frequency range $\omega/(2\pi) = $0.5-12 [GHz]. As one can see from the figure, for a  low frequency $\omega/(2\pi) < 7.2$ [GHz] the system is well-tuned, featuring a complete magnetization reversal triggered by the ferroelectric component of the structure. The increase in the  frequency results in a linear increase of the coercetivity field for the ferromagnetic hysteresis. The low-frequency hysteresis has a sharp and a well-defined shape (Fig. 4a). {The slow response of the FM to the faster oscillations of the FE polarization becomes noticeable under the increase of $\omega$, which manifests itself by rounded "corners" of the hysteresis curve (Fig. 4b) that becomes more pronounced for a frequency around 7 [GHz]. At this value, one can also observe a slight decrease in the magnetic remanence (Fig. 4c). When $\omega/(2\pi)$ exceeds the characteristic frequency of iron $2 \gamma K_1 / [(1+\alpha^2_{\mathrm{FM}})M_S  \mu_0] = 7.89$ [GHz], the behavior of the system changes drastically. The FM part does not follow the fast FE dynamics so that the curve of the magnetic hysteresis features several "breaks" and its coercitivity slightly drops down. It is essential to stress that, because we are dealing with a complex multi-site system driven by FE/FM interaction, the changes of the magnetization dynamics are noticeable for the frequency of electric fields slightly lower than the characteristic frequency of iron}. The magnetization reversal is accompanied with a strong deviation of $m_{\mathrm{y}}$ and $m_{\mathrm{z}}$ components (Fig. 4d). A further increase of the  frequency destabilizes the system even more, with the magnetization hysteresis looking ``skewed" and the coercitivity values ``jumping" back and forth  between two branches (Fig. 4e). It should be mentioned that the FM remanence lowers constantly for  a frequency $\omega/(2\pi) > 6.5$ [GHz], signaling that the achieved operation modes are not very promising for  applications. Also it is important to stress that the considered value of the coupling constant $\lambda = 12$ [s/F], while allowing a good influence of the FE over the FM layer, provides a minimal feedback that can be seen in a lack of the drastic changes of the remanence and the coercitivity for the FE part of the system. It goes without saying that such a pronounced unidirectional connection is an important feature for a device that is aimed to control the magnetic dynamics with an electric field {\it via} FE/FM coupling.\\
Based on these simulations, we would recommend to use the described compound multiferroic system under \textit{low} GHz frequency of the applied electric fields.

\begin{figure*}
\includegraphics[scale=1.2]{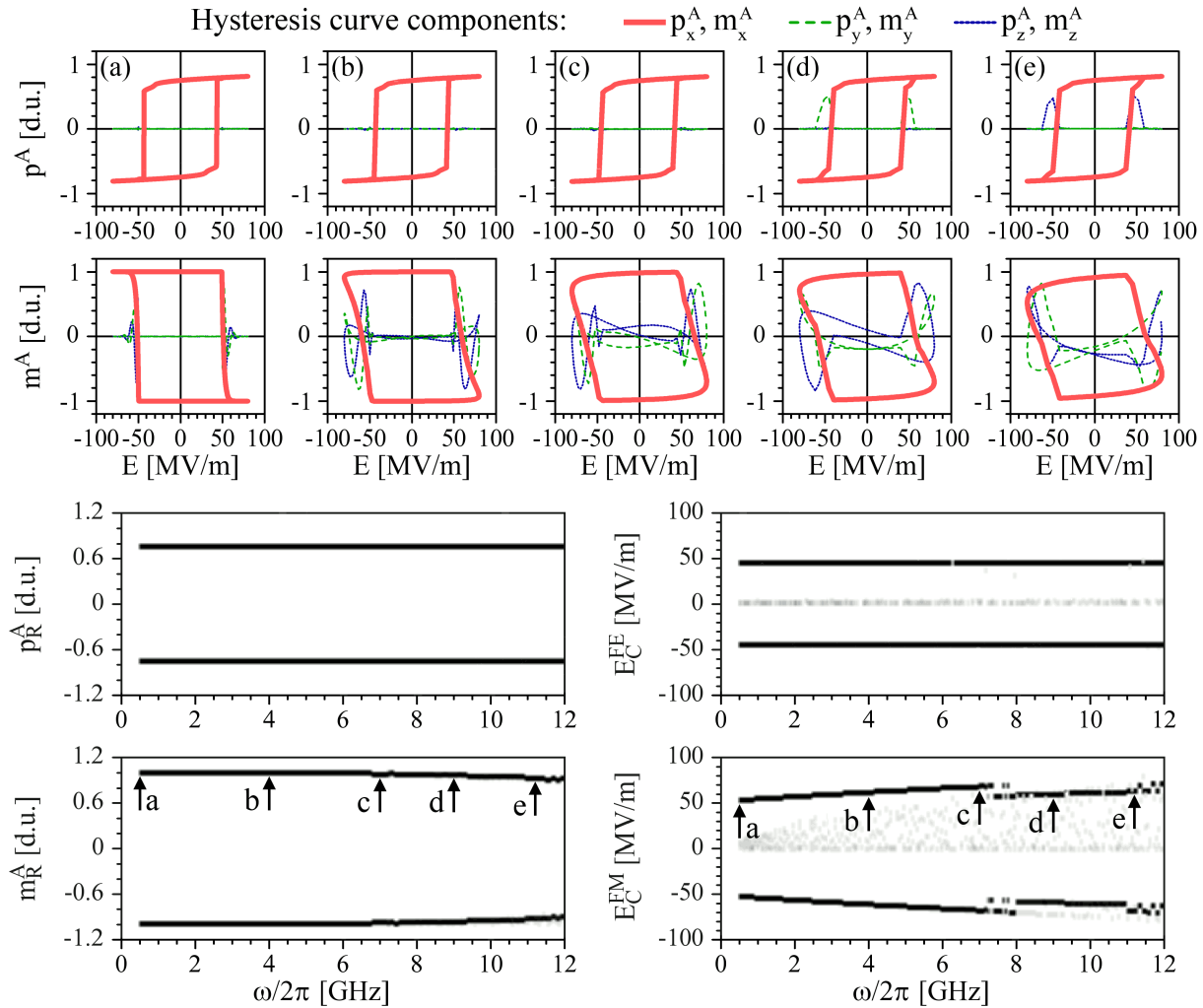}
\label{Fig4}
\caption{The frequency dependence of the FE/FM reversal for $\lambda = 12$ [s/F]. The most characteristic hysteresis curves are given for the following values of the field frequency $\omega/(2\pi)$: a) 0.5 [GHz], b) 4 [GHz], c) 7 [GHz], d) 9 [GHz] and e) 11.4 [GHz].}
\end{figure*}

\section{Conclusions}

We reported on a full-scale magnetization reversal in a composite multiferroic chain using an applied harmonic electric field with frequencies $0.5-12$ [GHz]. The dynamics and in particular the reversal   depend
 sensitively on the strength  of the magneto-electric coupling  $\lambda$. Hence, tracing the dynamics should
  deliver information on this coupling.
For a weak  coupling $\lambda < 0.8$ [s/F], the ferroelectric part shows well-defined hysteresis loop which
 may result in periodic oscillations of the ferromagnetic part.   A saturation magnetization $M_S$ may not be reached however. A magnetic hysteresis opens when $\lambda$ grows above 0.8 [s/F]. For a coupling constant strength  in the ranges $10-20$ [s/F], the multiferroic system has an optimal performance with both FE and FM hysteresis curves featuring a high remanence and a  low coercitivity. It is important to highlight that the range of $\lambda$ that corresponds to the FM hysteresis with the most definite shape does not vary with the frequency of the applied electric field. When the coupling constant becomes too strong (exceeding 40 [s/F]) the system does not show any hysteresis. This degradation is caused by the fact that the ferroelectric dynamics is strongly  disturbed by the ferromagnetic layer.

 We considered here one type of magnetoelectric coupling based on a screening model at the
 interface of the FE/FM parts. A different type of magnetoelectric coupling mechanism, e.g.
 a stress-strain coupling or a dynamical Dzyaloshinskii-Moriya coupling \cite{jia2011},
  may qualitatively alter the
 coupled dynamics  (even when the coupling strengths are comparable),  for the corresponding   functional forms
   entering the equation of motion are
 different in general.  Hence, in addition to the relevance for application, an attractive feature of studying the
 composite  dynamics is that it may deliver some details  on the underlying multiferroic coupling mechanisms.
 Studies along this line are currently underway.

\section{Acknowledgements}

The authors gratefully acknowledge M. Alexe and I. Vrejoiu for fuitful discussions. This work has been supported by the grants of CONACYT as Basic Science Projects 129269 and 133252 (Mexico), by the German Research Foundation SU 690/1-1 and by the SFB 762.


\begin{thebibliography}{99}
\bibitem{FM_damping}
The damping coefficient varies with the size of the computational cell. The value used in this paper was obtained by ``equalization" of the items appearing in the right hand side of equation (\ref{eq_13}), ensuring numerical stability of the system and a reasonable relaxation of the magnetization vector.
\end{thebibliography}
\end{document}